\documentclass[12pt]{article}
\begin{document}
\thispagestyle{empty}
\begin{flushright}
MPI-Ph/95-42\\
UNIGRZ-UTP-15-05-95\\
hep-th/9505092\\
Mai 1995
\end{flushright}
\vskip 1cm
\begin{center}
{\huge{On the U(1)-Problem of $\mbox{QED}_2$}}
\vskip 1cm
\centerline{ {\bf
Christof Gattringer${}^*$ }}
\vskip 5mm
\centerline{Max-Planck-Institut f\"{u}r
Physik, Werner-Heisenberg-Institut}
\centerline{F\"ohringer Ring 6, 80805 Munich, Germany}
\vskip 2mm
\centerline{and}
\vskip 2mm
\centerline{Institut f\"{u}r Theoretische Physik der Universit\"at Graz}
\centerline{Universit\"atsplatz 5, 8010 Graz, Austria}
\vskip3cm
\end{center}
\begin{abstract}
\noindent
$\mbox{QED}_2$ with mass and flavor has in common many features
with QCD, and thus is an interesting toy model relevant for four
dimensional physics. The model
is constructed using Euclidean path integrals and mass
perturbation series. The vacuum functional is carefully decomposed
into clustering states being the analogue of the $\theta$-vacuum of
QCD. Finally the clustering theory can be mapped onto a
generalized Sine-Gordon
model. Having at hand this bosonized version, several lessons on
the $\theta$-vacuum, the U(1)-problem and Witten-Veneziano-type
formulas will be drawn. This sheds light on the corresponding
structures of QCD.
\end{abstract}
\vskip 1cm
\bigskip \nopagebreak \begin{flushleft} \rule{2 in}{0.03cm}
\\ {\footnotesize \ Seminar talk given at
`Low-Dimensional Models in Statistical Physics and Quantum Field Theory',
34. Internationale Universit\"atswochen f\"ur Kern- und Teilchenphysik,
Schladming, Styria, Austria,
March 4-11, 1995 (to appear in the proceedings)
\vskip1mm
\noindent
${}^*$ e-mail: chg@mppmu.mpg.de}
\end{flushleft}
\newpage
%
%-----------------------------------------------------------
%
\section{Introduction}
$\mbox{QED}_2$ is a model which has many features in common with
(4-dimensional) QCD. U(1) gauge fields in two dimensions
allow for topologically nontrivial configurations which are the
counterparts of the instantons of 4-dimensional Yang-Mills theory. Thus
the formal linear combination of such sectors to the $\theta$-vacuum of
QCD \cite{callan} can be repeated for $\mbox{QED}_2$.
Furthermore the U(1)-axial
current acquires an anomaly and the corresponding pseudoscalar
particle turns
out to be massive as it is the case for QCD. The quoted features
allow to take over the formal reasoning that leads
to the formulation of the U(1)-problem \cite{u1}.

The purpose of this work is to construct $\mbox{QED}_2$ with mass
and flavor and study the announced topics ($\theta$-vacua, U(1)-problem)
{\it without} making use of poorly
defined concepts such as the formal superposition of topologically
nontrivial sectors to the $\theta$-vacuum or the definition of
non-gauge invariant currents (as it is done in the standard formulation
of the U(1)-problem).

Clustering $\theta$-vacua will be defined
using a mathematically rigorous limit procedure
which replaces the formal instanton arguments. The properties of this
vacuum state will be compared to the $\theta$-vacuum of QCD. Using the
method of bosonization we investigate the axial U(1)-symmetry on the
physical Hilbert space and discuss the status of the U(1)-problem.
Finally Witten-Veneziano-type formulas \cite{wv} that relate the masses of the
pseudoscalars to the topological susceptibility will be tested.

%
%-----------------------------------------------------------
%
\section{The model and techniques applied}
The Euclidean action of the model we consider is given by
$S = S_G + S_h +
S_F + S_M $.
The gauge field action reads
\begin{equation}
S_G[A] \; = \; \int d^2x\left[ \frac{1}{4} F_{\mu \nu}(x) F_{\mu \nu}(x)
+ \frac{1}{2} \lambda \Big( \partial_\mu A_\mu(x) \Big)^2 \right] \; .
\end{equation}
A gauge fixing term is included that will be considered in the limit
$\lambda \rightarrow \infty$ which ensures $\partial_\mu A_\mu = 0$
(transverse gauge). As usual, $F_{\mu \nu} =
\partial_\mu A_\nu - \partial_\nu A_\mu$, denotes the field strength
tensor.
The fermion action is a sum over N flavor degrees of freedom
\begin{equation}
S_F[\overline{\psi},\psi,A,h] \; = \; \sum_{f=1}^N \int d^2x \;
\overline{\psi}^{(f)}(x)\gamma_\mu \Big( \partial_\mu - i e A_\mu(x)
-i \sqrt{g} h_\mu(x) \Big)\psi^{(f)}(x) \; .
\end{equation}
Since the mass term will be treated perturbatively
we denote it separately
\begin{equation}
S_M[\overline{\psi}, \psi] \; = \; -\sum_{f=1}^N m^{(f)} \int d^2x
\; \chi_\Lambda(x) \;
\overline{\psi}^{(f)}(x) \psi^{(f)}(x) \; .
\end{equation}
$m^{(f)}$ are the fermion masses for the various flavors. In order
to prove the convergence of the mass perturbation series for more
than one flavor, the fields in
the mass term have to be cut off outside some finite rectangle $\Lambda$
in space-time. $\chi_\Lambda$ denotes the characteristic function
of $\Lambda$.

In addition to the gauge field an auxiliary vector field $h_\mu$
is coupled to the fermions (compare Equation (2)). Its action is given by
\begin{equation}
S_h[h] = \frac{1}{2}\int d^2x h_\mu(x)\Big( \delta_{\mu \nu} -
\lambda^\prime \partial_\mu \partial_\nu \Big) h_\nu(x) \; .
\end{equation}
$S_h[h]$ is simply a white noise term plus a term that
makes $h_\mu$ transverse in the limit $\lambda^\prime \rightarrow \infty$.
When integrating out $h_\mu$ it generates a Thirring term
for the U(N) flavor singlet current. The purpose of this Thirring
term is to make the short distance singularity of
$\overline{\psi}^{(f)}(x) {\psi}^{(f)}(x) \;
\overline{\psi}^{(f)}(y) {\psi}^{(f)}(y) $
integrable. This expression is a typical term showing up in
a power series expansion of the mass term (3). It has to be integrated over
$d^2x d^2y$ which is possible only if an ultraviolet regulator
such as the Thirring term is included.

The approach we adopted is the quantization via Euclidean functional
integrals. After having expanded the exponential of the mass term of
the action $\exp(-S_M[\overline{\psi},\psi])$, vacuum
expectation values of operators ${\cal O}$ are given by
\begin{equation}
\langle {\cal O}[\overline{\psi},\psi,A,h] \rangle \; = \;
\frac{1}{Z} \sum_{n=0}^\infty
\frac{(-1)^n}{n!}
\langle {\cal O} [\overline{\psi},\psi,A,h]
\Big(S_M[\overline{\psi},\psi]\Big)^n \rangle_0 \; ,
\end{equation}
where the expectation values of the massless model are formally defined as
$
\langle {\cal P}  \rangle_0 \; = \;
Z_0^{-1} \int {\cal D}h {\cal D}A {\cal D}\overline{\psi} {\cal D}\psi \;
{\cal P} \;
e^{ - S_G -S_h -S_F} \; .
$
Of course also the normalization constant $Z$ showing up in (5) has to
be expanded in terms of the massless theory.

When giving a precise mathematical meaning to the so far poorly defined
functional integral one usually starts with integrating out the fermions.
This gives rise to the fermion determinant, which can be constructed
properly when the fermions are massless
(this is the reason why the mass term is treated perturbatively).
The renormalized determinant can be found in e.g. \cite{seiler} and reads
(the potentials $A_\mu, h_\mu$ are assumed to be transverse and
to satisfy some mild regularity and falloff
conditions at infinity)
$
\mbox{det}_{ren} [ A_\mu, h_\mu] \; = \;
\exp\big( (2\pi)^{-1} \| e A^T + \sqrt{g}h^T \|_2^2 \big) \; ,
$
where the superscript $T$ denotes the transverse part of the vector
fields. Adding the exponent of this expression
to the action one ends up with
the effective action for the gauge field and the auxiliary field.
Both those terms are Gaussian and the vacuum expectation values
obtain a
precise mathematical meaning through
Gaussian functional integrals. Since the dependence of the propagators
on the external fields is exponential the functional integrals
can be computed.

The vacuum expectation functional constructed for the massless model
so far violates the cluster
decomposition property. In particular it turns out (see e.g. \cite{gattringer})
that operators which are singlets
under $\mbox{U(1)}_V \times \mbox{SU(N)}_L\times \mbox{SU(N)}_R$, but
transform nontrivially under $\mbox{U(1)}_A$, do not cluster.
This implies that the vacuum state is not unique. In order to
obtain a proper vacuum, we use a limit process \cite{gattringer}.
The original operator
${\cal P}$ is correlated with a test operator
${\cal U}_\tau ({\scriptstyle{\cal P}})$ depending on the chiral charge
$Q_5({\scriptstyle{\cal P}})$ of ${\cal P}$
and then the limit of shifting the test operator
to timelike infinity is considered.
\begin{equation}
\langle {\cal P}({\scriptstyle\{x\}}) \rangle^\theta_0 \; := \;
e^{i \theta \frac{Q_5({\scriptscriptstyle{\cal P}})}{2N}}
\lim_{\tau \rightarrow \infty}
\langle {\cal U}_\tau ({\scriptstyle {\cal P}}) \;
{\cal P}({\scriptstyle\{x\}}) \rangle_0 \; .
\end{equation}
The set of `test operators' ${\cal U}_\tau ({\scriptstyle {\cal B}})$
is defined by
\begin{equation}
{\cal U}_\tau ({\scriptstyle {\cal P}}) \; := \;
\left\{ \begin{array}{l}
{\cal N}^{(n)} (\{y\})  \prod_{i=1}^n {\cal O}_\mp (\{y+\hat{\tau}\})
\; \mbox{for} \; \; Q_5({\scriptstyle{\cal P}}) = \pm 2nN , \;
n \geq 1 \; , \\ \; \\
1 \; \; \mbox{otherwise} \; \; , \end{array} \right.
\end{equation}
where
${\cal O}_\pm ( \{y\} ) :=
\prod_{f=1}^{N}\overline{\psi}^{(f)}(y^{(f)})
\; \frac{1}{2}[1 \pm \gamma_5] \; \psi^{(f)}(y^{(f)})$.
The space-time arguments $y^{(f)}$
are arbitrary and the left hand side of (7)
does not depend on them due to a normalizing factor ${\cal N}^{(n)}(\{y\})$.
$\theta$ is a real parameter in the range $[0,2\pi)$.
For this decomposition prescription the following theorem was proven
\cite{gattringer}.
\vskip1mm
\noindent
{\bf Theorem 1.}\\
{\bf i) } {\it The cluster decomposition property holds for}
$\langle .. \rangle^\theta_0$.\\
{\bf ii) } {\it The state $\langle .. \rangle_0$ constructed initially
is recovered by averaging over $\theta$, and thus is a mixture of the pure
states $\langle .. \rangle_0^\theta$.}
\vskip2mm
\noindent
The decomposition (6) replaces the mathematically poorly
defined concept of superimposing topological sectors to the $\theta$-vacuum
\cite{callan}.

In two dimensions one has at hand the powerful tool of
bosonization. It will be used to construct the massive model.
By evaluating a generating functional including vector currents
and chiral densities one can establish the following
bosonization prescriptions
\begin{equation}
J^{(I)}_\nu(x) \; \longleftrightarrow \;
\left\{ \begin{array}{ll}
- \frac{1}{\sqrt{\pi + gN}} \;
\varepsilon_{\mu \nu} \partial_\mu \Phi^{(1)}(x) \; \; \; \; \; \; & I = 1 \\
\; & \;   \\
- \frac{1}{\sqrt{\pi}} \;
\varepsilon_{\mu \nu} \partial_\mu \Phi^{(I)}(x) \; \; \; \; \; \;
& I = 2,\; ... \; N \; .
\end{array} \right.
\end{equation}
The currents are defined as $J_\mu^{(I)} := \sum_{f,f^\prime=1}^N \;
\overline{\psi}^{(f)} \gamma_\mu H^{(I)}_{ff^\prime} \; \psi^{(f^\prime)}$.
The mixing matrices $H^{(I)}$ in flavor space are chosen to be proportional
to the unit matrix for $H^{(1)}$, and proportional to the generators of a
Cartan subalgebra of SU(N) for $H^{(I)}, \; I = 2,3, .. N$
(see \cite{gattringer} for the explicit definition).
The chiral densities are bosonized by
\[
\overline{\psi}^{(f)}(x) \; \frac{1}{2}
[1 \pm \gamma_5 ] \psi^{(f)}(x) \; \longleftrightarrow
\]
\vspace{-3mm}
\begin{equation}
\frac{1}{2\pi} c^{(f)}
: e^{\mp i 2 \sqrt{\pi} \sqrt{\frac{\pi}{\pi +gN}} U_{1 f} \Phi^{(1)}(x)}
:_{M^{(1)}} \; \prod_{I=2}^N
: e^{\mp i 2 \sqrt{\pi} U_{I f} \Phi^{(I)}(x)}
:_{M^{(I)}} e^{\pm i \frac{\theta}{N}}\; .
\end{equation}
$U_{I f}$ are some real factors that are related to the entries of
the matrices $H^{(I)}$. Normal ordering of the exponentials
with respect to some reference mass $M$ (see e.g. \cite{frohlich})
is denoted by $: .. :_M$.

The Lagrangian of the bosonic theory was obtained by bosonizing the
terms of the mass expansion (5) and then summing up the perturbation series.
One ends up with the following generalized Sine Gordon model
\[
{\cal L}_{GSG} \; = \;
\frac{1}{2} \sum_{I=1}^N \partial_\mu \Phi^{(I)} \partial_\mu \Phi^{(I)}
\; + \; \frac{1}{2} \Big( \Phi^{(1)} \Big)^2 \; \frac{e^2 N}{\pi+gN}
\]
\vspace{-3mm}
\begin{equation}
- \; \frac{1}{\pi} \sum_{f=1}^N m^{(f)} c^{(f)}
: \cos \left(
2 \frac{\pi}{\sqrt{\pi\!+\!gN}} U_{1 f} \Phi^{(1)}
+ 2\sqrt{\pi} \sum_{I=2}^N U_{I f} \Phi^{(I)} - \frac{\theta}{N} \right):
{}.
\end{equation}
Normal ordering of the cosine is understood in the sense of the
mass-expansion and thus reduces to the normal ordering of exponentials.
It has to be stressed that this normal ordering is essential
for the correct bosonization of the clustering states
$\langle .. \rangle^\theta$. In particular the normal ordered
exponentials of the massless fields $\Phi^{(I)}, I \geq 2$ have to
obey the neutrality condition \cite{frohlich}, to give nonvanishing
expectation values. This ensures the correct assignment of
expectation values to operators with nonvanishing chiral charge.
A model similar to (10) but without Thirring term
was discussed in \cite{belvedere} using an operator approach.

Using the bosonized version of the model we were able to prove the
following theorem \cite{gattringer}
that generalizes the
N=1 proof by Fr\"ohlich \cite{frohlich}.
\vskip1mm
\noindent
{\bf Theorem 2.} \\
{\it The mass perturbation series converges for finite $\Lambda$ and
$m^{(b)} < r(\Lambda)$.}
\vskip2mm
\noindent
Due to the massless degrees of freedom the applied techniques only
allowed to establish a
radius of convergence $r(\Lambda)$ which shrinks to zero when sending
$\Lambda$ to infinity. We believe that the proof can be made better
(i.e. $r$ becomes independent from $\Lambda$) by
applying some generalized cluster expansion methods.

This concludes the construction of the model and one can start to
discuss its physical implications.

%
%-----------------------------------------------------------
%
\section{Lessons on the U(1)-problem}
The following three lessons formulate the physically
interesting results that were obtained for $\mbox{QED}_2$, and shed
light on the corresponding problems in $QCD$.
\\
\noindent
{\bf Lesson 1 :}
{\it The structure
of the vacuum functional that has been suggested
within the instanton picture is recovered from the mathematically rigorous
construction given in formula} (6).
\\
\noindent
In particular only operators with chirality
$2 \mbox{N} \nu \; , \; \nu \in
\mbox{Z\hspace{-1.35mm}Z}$ have nonvanishing vacuum expectation values.
This can be seen immediately from the prescription (6) for the $\theta$-vacuum.
The same has been claimed by 't Hooft for QCD \cite{thooft}. Furthermore
an analysis of the Lagrangian (10) shows that as long as one of the fermion
masses $m^{(f)}$ vanishes physics is independent of $\theta$
(see \cite{callan} for the case of QCD).
\vskip1mm
\noindent
{\bf Lesson 2 :}
{\it The axial U(1)-symmetry is not a symmetry on the physical
Hilbert space, and there is no U(1)-problem for $QED_2$.}
\\
\noindent
Under a U(1)-axial transformation all the chiral densities
$\overline{\psi}^{(f)}
(1 \pm \gamma_5 )/2 \psi^{(f)}$ obtain phases $\exp( \pm i2\varepsilon)$.
This transformation property is not compatible
with the structure of the coefficients $U_{I f}$ that show up in
the bosonization prescription (9) (see \cite{gattringer} for the explicit
proof). Thus even in the case of
vanishing fermion masses $m^{(f)}$ the axial U(1)-symmetry is not realized
on the physical Hilbert space, and thus the U(1)-problem is not there at all.
The same could be true for QCD, since the current that is used to construct
the axial symmetry is not gauge invariant.
\vskip1mm
\noindent
{\bf Lesson 3 :}
{\it The masses of the particles that correspond to the vector currents
obey a Witten-Veneziano type formula.}
\\
\noindent
This result was established using a semiclassical approximation of
the Lagrangian (10). The approximation is necessary due to the
problem with removing the cutoff $\Lambda$. Nevertheless this
approximation is under good control, since for small fermion
masses $m^{(f)}$ it reduces to the exact solution.
%
%-----------------------------------------------------------
%

\end{document}